\documentclass[conference]{IEEEtran}
\usepackage[T1]{fontenc}
\usepackage{amsmath,amssymb,amsfonts}
\usepackage{booktabs,cite,graphicx,microtype,url}
\usepackage[hidelinks]{hyperref}
\usepackage{xcolor}
\newif\ifshowrev
\ifdefined\showredmarks \showrevtrue \else \showrevfalse \fi
\ifshowrev
  \newcommand{\rev}[1]{{\color{red}#1}}
\else
  \newcommand{\rev}[1]{#1}
\fi
\newcommand{\QOriginal}{185}
\newcommand{\QCombined}{196}
\newcommand{\QSharedRank}{193}
\newcommand{\QMulti}{190}
\newcommand{\PrimaryDen}{196}
\newcommand{\NWDen}{42}
\newcommand{\CalPrimary}{147}
\newcommand{\CalNW}{28}

\newcommand{\QLowOriginal}{173}
\newcommand{\QLowCombined}{194}
\newcommand{\ExampleA}{9}
\newcommand{\ExampleH}{4}
\newcommand{\ExampleG}{0.13001765}
\newcommand{\ExampleStaticA}{12}

\newcommand{\FreshFits}{900{,}000}

\newcommand{\FreshStates}{30}
\newcommand{\FreshAliases}{204}
\newcommand{\FreshFamily}{300}
\newcommand{\FreshModelError}{1.32}

\newcommand{\FreshFprMin}{4.62}
\newcommand{\FreshFprMax}{5.62}

\newcommand{\FreshRefG}{12.84}
\newcommand{\FreshRefLo}{11.12}
\newcommand{\FreshRefHi}{14.70}

\newcommand{\FreshRefSecondaryLo}{10.62}
\newcommand{\FreshRefSecondaryHi}{14.13}

\newcommand{\FreshQG}{13.42}
\newcommand{\FreshQLo}{11.67}
\newcommand{\FreshQHi}{15.32}

\newcommand{\FreshNWLo}{8.71}
\newcommand{\FreshNWHi}{11.96}

\newcommand{\FreshNWSecondaryLo}{8.47}
\newcommand{\FreshNWSecondaryHi}{11.69}

\newcommand{\RPrimaryKthreeSfour}{13.3108}
\newcommand{\RPrimaryKfourSfour}{14.0657}
\newcommand{\RPrimaryKtwoSfour}{10.5116}
\newcommand{\RNWOneKsixteenSfour}{11.1346}
\newcommand{\RNWOneKfourSfour}{10.7431}

\newcommand{\DirectLow}{188}

\newcommand{\XK}{\mathcal{X}_{K}}
\newcommand{\one}{\mathbf{1}}
\title{Quantum-Assisted D-FACTS Scheduling for Moving Target Defense under Switching Limits}
\author{\IEEEauthorblockN{Hongyu Wu\textsuperscript{1}, Yichen Liu\textsuperscript{1}, and Bo Liu\textsuperscript{2}}
\IEEEauthorblockA{\textsuperscript{1}Mike Wiegers Department of Electrical and Computer Engineering, Kansas State University, Manhattan, United States\\
\textsuperscript{2}School of Engineering and Applied Sciences, Washington State University, Richland, United States}}
\begin{document}
\bstctlcite{IEEEexample:BSTcontrol}
\maketitle

\begin{abstract}

Moving target defense with distributed flexible alternating-current transmission
system (D-FACTS) devices requires activation decisions as operating points
change. Switching limits couple these decisions, but period-by-period rankings
can overlook combinations needed to complete low-activation schedules.
This paper proposes cross-period candidate sharing and schedule-completion
evaluation to allocate limited physical checks in quantum-assisted D-FACTS
scheduling. The quantum approximate optimization algorithm (QAOA)
generates activation candidates, while classical scheduling enforces operating
and switching constraints. We also quantify model-based activation requirements
and test detection through noisy nonlinear trials. A modified IEEE 118-bus
study assesses recovery of reference activation and switching objectives within
the evaluated set. Combining candidate sharing and schedule-completion
evaluation improves final-budget recovery over period-by-period ranking in the
primary operating window for every tested candidate generator. The results show that
quantum and classical candidates yield identical recovery curves under combined checking, so the same allocation policy applies to either source.

\end{abstract}

\section{Introduction}

An attacker who knows the \rev{relationship} between \rev{system} states and measurements
can construct \rev{false data injection attacks that evade residual-based detection.}
Moving target defense (MTD) \rev{perturbs this relationship by changing line
reactances through} distributed flexible
AC transmission system (D-FACTS) \rev{devices}
\cite{Zhang2022Voltage}. \rev{Consequently, attacks constructed with pre-MTD
parameters may become detectable. However, reactance perturbations also affect
power flows and bus voltages, so MTD decisions need to account for both detection
effectiveness and operating constraints.}

MTD \rev{studies address} D-FACTS placement and operation.
Placement studies \rev{improve detection coverage}
\cite{LiuWu2020Placement,LiuWu2021Systematic}, \rev{while cost-benefit analysis
\cite{Lakshminarayana2021Cost} and voltage-stability-constrained MTD
\cite{Zhang2022Voltage} examine the costs and operating effects of reactance
perturbations. These studies do not directly address activation scheduling
across operating points.}

\rev{Operational MTD studies select defense actions over time. Game-theoretic
MTD learns randomized line-selection policies \cite{Lakshminarayana2021Game},
while multistage MTD sequences security- and economy-oriented perturbations
\cite{Wang2023MMTD}. Event-triggered MTD uses alarms to guide defense
\cite{Xu2023EventTriggered}, and robust MTD considers worst-case attacks
\cite{Xu2023Robust}. These approaches address defense selection rather than
allocating limited physical evaluations among candidates.}

With \rev{device} locations and reactance settings fixed, switching limits couple
activation \rev{decisions across periods.
Evaluating each period-combination pair} requires AC optimal power flow, continuation
power flow, and detector calculations\rev{. Because a combination's value also
depends on its compatibility with adjacent periods, period-by-period rankings may}
overlook combinations needed \rev{for feasible,} low-activation schedules.
\rev{We therefore propose} cross-period candidate sharing and
schedule-completion evaluation \rev{to allocate limited physical checks. The
scheduling objective minimizes device-periods (active devices summed over
periods) under model-based detection and switching constraints, with exhaustive
evaluation providing reference objectives within the evaluated set.}

\rev{We use the} quantum approximate optimization algorithm (QAOA) \rev{to generate
fixed-count D-FACTS activation candidates for subsequent physical evaluation.
Each device is represented by one qubit, and a} fitted quadratic detection
response defines the cost operator.
Cost-dependent phases and XY mixing \rev{guide candidate sampling,} while ideal
XY operations preserve the activation count \rev{without requiring a count penalty}
\cite{Hadfield2019Alternating,Wang2020XYMixers}.
\rev{Physical evaluation and schedule optimization remain classical, so
quantum candidate generation is kept separate from the enforcement of operating and switching
constraints.}

\rev{Prior power-system QAOA studies have explored parameter transfer and
constraint-aware circuit design.} Jing et al.\ transfer circuit parameters between weighted
graphs to reduce parameter search \cite{Jing2023DataDriven}\rev{, while the}
PACE-QAOA preprint uses physical constraints
to reduce \rev{encoding requirements for controlled islanding} \cite{Jiang2026PACE}.
\rev{In this work, QAOA generates D-FACTS activation candidates for
schedule-aware physical evaluation. We assess whether these} candidates recover reference
activation and switching objectives \rev{under limited physical evaluations.
Comparisons with five classical candidate generators, using the same fitted responses and
evaluation budgets, examine the effect of candidate generation, while
fixed-candidate comparisons isolate the effect of checking policies.} The main contributions of this paper are as follows.
\begin{itemize}
\item We propose \rev{a schedule-aware physical-evaluation strategy combining}
cross-period candidate sharing and schedule-completion evaluation.
\rev{Candidate sharing expands the combinations available to each period, while
schedule completion prioritizes checks that can complete switching-feasible schedules.
Their combined use, termed combined checking, improves final-budget recovery of
reference activation and switching objectives} over period-by-period ranking in
the primary window for every tested \rev{candidate} generator.
\item We \rev{characterize} model-based activation requirements across detection
\rev{requirements}, switching allowances, and operating windows.
\rev{Noisy nonlinear state-estimation trials distinguish recovery of scheduling
objectives from empirical confirmation of detection requirements.} \rev{In one
tested IEEE 118-bus window, switching reduces device-periods without reducing}
the number of distinct devices used.
\end{itemize}

\section{Operating and Attack Models}
\subsection{AC operation and loading margin}
We schedule 16 candidate devices from the installed D-FACTS set.
For \(x\in\{0,1\}^{16}\), \(x_i=1\) activates device \(i\)
at fixed reactance multiplier \(1+\rho_i\).
AC optimal power flow (OPF) redispatches each period-combination case.
Acceptance requires
\begin{equation}
 h_t^{\rm AC}(x)=1,\qquad \lambda_t(x)-\lambda_t^0\geq-0.01,
 \label{eq:accepted}
\end{equation}
where \(h_t^{\rm AC}\) requires convergence and bus voltages within
0.94-1.06 p.u. The base and post-MTD continuation power flow (CPF)
loading parameters are \(\lambda_t^0\) and \(\lambda_t(x)\), respectively.

CPF uses the reactance-modified input case rather than the redispatched
OPF solution. Its target doubles nonzero PQ-bus demands, and it traces the
load increase to the nose without finite operating-event limits.
Thus, \eqref{eq:accepted} constrains the input-case nose parameter rather than
the margin after redispatch. 
The steady-state model excludes
generator ramping and switching transients.

\subsection{Measurements and attacker knowledge}
Post-MTD measurements comprise bus-voltage magnitudes, active and reactive
injections, and flows in both directions on 179 branches of the IEEE 118-bus system.
Fixing one angle leaves 235 state variables for 1070 measurements.
For state \(\theta_t\) and MTD combination \(x\), the estimator uses
\begin{equation}
 z_t(x)=h_t(\theta_t,x)+e_t,\qquad
 e_t\sim\mathcal N(0,\Sigma_t(x;\sigma)).
 \label{eq:measurement}
\end{equation}
Power measurements are normalized by the case base MVA. Diagonal covariance entries are
\(\sigma^2h_{t,j}^2\), with values below \(10^{-20}\) replaced by \(5\times10^{-9}\).
Clean tables use OPF generation-minus-load injections and branch flows.
Their numerical mismatch with the estimator map is retained.

At each no-MTD point, a net load redistribution attack (NLRA) replaces
voltages and injections at buses 82-112 and internal directed flows.
Boundary-crossing flows remain unchanged. Nominal replacement and clean tables
define the context-\(s\) increment
\begin{equation}
 \Delta z^0_{t,s}=z^{0,\rm rep}_{t,s}-z_t^{\rm base},\qquad
 \widetilde z_{t,s}(x)=z_t^{\rm clean}(x)+\Delta z^0_{t,s}.
 \label{eq:attack}
\end{equation}
The attacker knows the nominal point but not the realized activation.
The increment stays fixed after reactance changes and redispatch, so the
post-MTD attack is additive. Its physical realizability after MTD and protection
against an attacker who updates the increment remain untested.

\subsection{Scheduling score and measured detection}
The weighted-least-squares (WLS) estimator fits voltage magnitudes and angles
locally. The detector statistic approximates the minimum residual
\begin{equation}
 J_t(z,x)=\min_\theta[z-h_t(\theta,x)]^{\mathsf T}
 \Sigma_t(x;\sigma)^{-1}[z-h_t(\theta,x)]\rev{.}
 \label{eq:wls}
\end{equation}
\rev{For each attack context, n}oiseless attacked measurements give reference residuals
\(J^{\rm ref}_{t,s}(x)\) at covariance noise 0.03. At target noise
\(\sigma=0.003\), \rev{the residual scales by \((0.03/0.003)^2=100\), which enters
the scheduling score as a noncentrality parameter:}
\begin{equation}
 P^{\rm model}_{D,t,s}(x)=
 \Pr\{\chi^2_{835}(100J^{\rm ref}_{t,s}(x))>\tau\},
 \label{eq:pd}
\end{equation}
where \(\tau=\chi^2_{0.95}(835)=903.3357\) corresponds to a nominal
5\% false-alarm probability. Because covariance replacement prevents exact scalar scaling,
\eqref{eq:pd} \rev{is approximate, so e}mpirical detection is assessed separately through noisy WLS trials. Neither this score nor
a detector alarm establishes the physical damage of an attack.

\section{Quantum Proposals and Schedule-Based Checks}
\subsection{Activation and switching decisions}
For \(T=3\), schedule \(X=(x_1,x_2,x_3)\), and declared attack
trajectories \(\omega=(s_1,s_2,s_3)\in\Omega\), we define minimum scheduling
score \(G\), device-period total \(A\), and on/off count \(H\) by
\begin{align}
 G(X)&=\min_{t,\omega\in\Omega}P^{\rm model}_{D,t,s_t}(x_t),
 \label{eq:worst}\\
 A(X)&=\sum_{t=1}^T\one^{\mathsf T}x_t,\qquad
 H(X)=\sum_{t=2}^T\|x_t-x_{t-1}\|_1.
 \label{eq:cost}
\end{align}
A replacement contributes two changes to \(H\).
Initial activation is excluded because its preceding state is unspecified.
With four contexts per period, \(G\) is the minimum of twelve probabilities,
not cumulative detection.

Let \(\XK=\{x:\one^{\mathsf T}x=K\}\). The evaluated set
\(\mathcal D=\bigcup_{K\in\{0,1,2,3,4,16\}}\XK\) has 2518 combinations.
For a checked set \(\mathcal C_t\subseteq\mathcal D\), let
\(\mathcal A_t(\mathcal C_t)\) represent the combinations accepted by
\eqref{eq:accepted}. Given detection requirement \(\eta\) and switching allowance
\(S\), the scheduler solves
\begin{equation}
 \begin{aligned}
 &\underset{x_t\in\mathcal A_t(\mathcal C_t)}{\operatorname{lexmin}}
       \ (A(X),H(X))\\
 &\text{subject to}\quad G(X)\geq\eta,\quad H(X)\leq S.
 \end{aligned}
 \label{eq:decision}
\end{equation}
Dynamic programming minimizes switching at each terminal combination and
activation total. The reference uses \(\mathcal C_t=\mathcal D\)\rev{.} Recovery means matching reference \((A,H)\),
not branch choices or confirmed detection.

\subsection{Candidate generation}
\rev{To rank candidates without physical checks, w}e
fit eight trajectory-indexed quadratic responses per period to all 137
combinations with at most two devices and 24 uniform samples each from
\(\mathcal X_3,\mathcal X_4\). The 185 fitting combinations exclude
the checked all-active case. Each response is
\begin{equation}
 \widehat F_{t,\omega}(x)=c_0+\sum_i c_i x_i+
                       \sum_{i<j}c_{ij}x_ix_j,
 \label{eq:fit}
\end{equation}
with target \(P^{\rm model}_{D,t,s_t}\), 137 features, and least-squares
cutoff \(10^{-12}\). Unclipped predictions determine the checking order,
whereas \rev{physically} evaluated scores determine detection feasibility.

Ideal QAOA circuits use \(K=3,4\). Let \(C_{t,\omega}\) be diagonal
with entries \(-\widehat F_{t,\omega}+c_0\), divided by the largest absolute
linear or pair coefficient, or one if all vanish.
One layer acts on the uniform \(K\)-active superposition \(|D_K\rangle\) as
\begin{equation}
 |\psi\rangle=U_{\rm XY}^{\rm ord}(\beta)
       e^{-i\gamma C_{t,\omega}}|D_K\rangle.
 \label{eq:qaoa}
\end{equation}
The ordered product \(U_{\rm XY}^{\rm ord}\) applies
\(\exp[-i\beta(X_iX_j+Y_iY_j)/2]\), with Pauli operators \(X_i,Y_i\),
in order \((1,2),(1,16),(2,3),\ldots,(15,16)\).
It preserves activation count but differs from summed-ring evolution.
Differential evolution optimizes
\(\gamma,\beta\in[0,2\pi]\) with exact expectations, at most 40 generations,
population multiplier eight, and polishing. Simulation enumerates
560- and 1820-state bases. Per count and period, four targets each use two
optimization/sampling seeds and 256 samples per instance.
Samples are pooled and deduplicated.

Classical controls use uniform sampling, aggregate-response and
scenario-conditioned swap searches, simulated annealing, and full quadratic
ranking with the same fits and checking rules.
Uniform and swap methods use 2048 draws or proposals per count.
Swap searches retain non-improving proposals. Annealing uses sixteen chains of
sixteen occupied states per scenario. Full ranking evaluates all 2380
three- and four-device predictions, not physical values.

\subsection{Candidate sharing and schedule completion}
Period-by-period ranking orders candidates by
\(\min_\omega\widehat F_{t,\omega}\), with identifiers for ties.
\rev{However, a candidate generated for one period may also help complete
schedules in other periods.}
Cross-period sharing \rev{therefore} pools candidates while predictions and evaluations
remain period-specific.

\rev{Schedule completion prioritizes unchecked combinations by their potential
to complete switching-feasible schedules. It} predicts up to three \rev{such}
schedules per requirement-cap pair. Successive searches exclude earlier unchecked
period-combination pairs. Unchecked pairs use predicted scores and provisional
acceptance, whereas operating schedules use only checked combinations.
Each five-check increment per count and period reserves checks for the two highest-ranked
remaining combinations. Further checks maximize predicted schedule completions
per added evaluation, summed over all requirement--cap pairs.
Ties use check count, \(A,H\), and identifiers. Unused checks follow ranking.
This heuristic need not recover the full-set optimum.

\rev{Frequency-ranking controls instead rank candidates by how often they
appear in predicted schedules, using}
one predicted schedule per requirement--cap pair
(direct temporal ranking) or combined checking's predicted schedules (the common bank).
Within each period and count, \rev{they order} candidates by decreasing
requirement--cap frequency, decreasing predicted minimum score, then increasing ID.
A second common-bank control reserves two priority checks per increment.
Later choices can differ. \rev{All controls share candidates and initial
evaluations and use} fixed orderings without retraining.

Budgets apply per window, candidate generator, treatment, and seed.
Requirement--cap pairs reuse evaluations. The initial \(186\times3=558\) evaluations grow to
588--738 after adding 5--30 combinations per count and period.
Equal evaluation budgets
do not imply equal computation, state preparation, sampling, or runtime.

\begin{figure*}[t]
 \centering
 \includegraphics[width=0.92\textwidth]{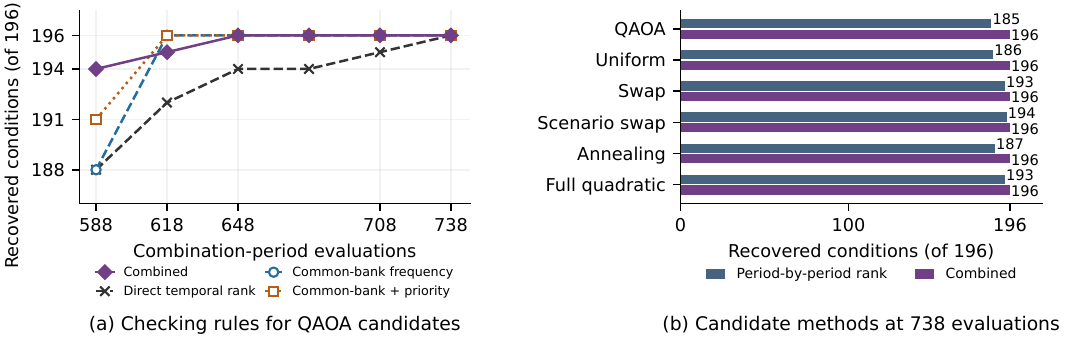}
 \caption{Reference-objective recovery for (a) checking policies with QAOA
 candidates and (b) candidate generators at the final budget.}
 \label{fig:budget}
\end{figure*}

\begin{figure}[t]
 \centering
 \includegraphics[width=\columnwidth]{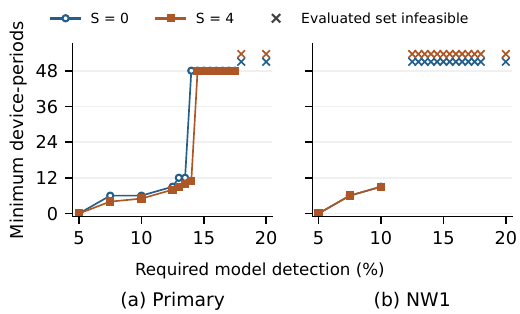}
 \caption{Minimum device-periods required at switching caps zero and four
 in the two operating windows.}
 \label{fig:activation}
\end{figure}

\begin{figure*}[t]
 \centering
 \includegraphics[width=0.86\textwidth]{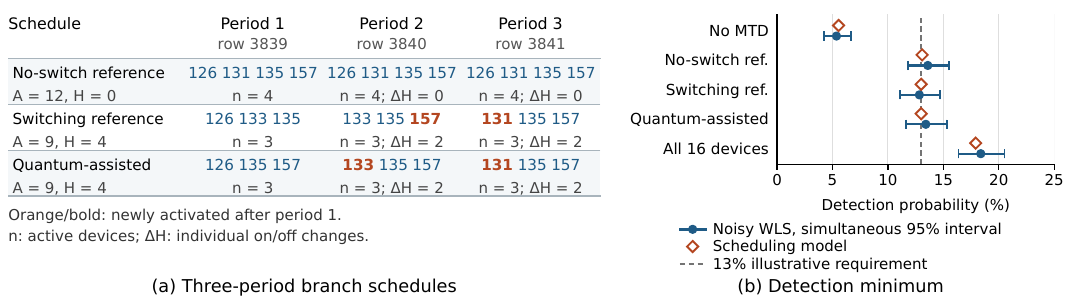}
 \caption{Branch schedules and minimum detection probabilities with simultaneous
 95\% confidence intervals in the primary window.}
 \label{fig:operation}
\end{figure*}

\section{Case Study}
\subsection{Windows, scenarios, and reference}
AC OPF, state estimation, and CPF follow \cite{Zhang2022Voltage} on the modified
IEEE 118-bus case \texttt{case118DCOPF\_Servor}.
Zero TAP inputs in OPF and CPF set nine nonunit transformer ratios to unity.
Of 62 installed D-FACTS devices, the 16 incident to buses 82--112 are scheduling
candidates. Seed 11 fixes their reactance changes between \(-12.67\%\) and
\(+11.49\%\). The other 46 retain nominal reactance.
All-active means these 16 devices, not all 62.

The primary window comprises rows 3839--3841, historically selected for maximum
adjacent variation in 91 load-scale columns. NW1 uses previously accessed
rows 1--3.
Attack contexts \(a,b,c,d\) target buses 98, 84, 93, and 86 at normalized
NLRA magnitude 0.2. Nominal attacks fix original generation, allow virtual
load injections within \(\pm20\%\), and retain source boundary constraints.
Trajectory order is \(aaa,bbb,abb,baa,ccc,ddd,cdd,dcc\).
All enter fitting and scoring, so none is unseen.

OPF/NLRA uses \(10^{-8}\) tolerances, a 150-iteration MIPS cap,
fresh nominal anchors, and complete attacks. WLS uses step tolerance
\(10^{-6}\), cap 1000, and the source pre-update residual.
The 15,108 evaluations reuse six no-MTD cases. All 75,540 reference WLS fits
\rev{converge with finite returns}.
All combinations pass AC/CPF, so acceptance does not explain method
differences. Reactance and redispatch effects are not separated.

The 18 requirements are 5\%, 7.5\%, 10\%, 12.5\%--18\% at 0.5-point
intervals, 20\%, 50\%, and 90\%. Caps \(S=0,4\) and paired seeds
101, 211, 307, 401, 503, 601, and 709 give 252 correlated conditions
per window and treatment.
Mixed-integer linear programming (MILP) independently verifies objective values for
12,096 final-quota and 72 reference decisions for the original four treatments.
Earlier quotas use dynamic programming. A second program checks differing
common-bank objective pairs.
The offline reference is outside each method's budget.

\subsection{Recovery under the evaluation budget}
Figure~\ref{fig:budget} compares recovery over \PrimaryDen{} reference-feasible
primary conditions (twenty-eight requirement-cap pairs per seed) and \NWDen{} in NW1.
Initial and fallback recovery is \CalPrimary{}/\PrimaryDen{} and
\CalNW{}/\NWDen{}, respectively.
Panel (a) holds QAOA candidates fixed.
At 738 evaluations, QAOA candidates recover \QOriginal{} primary-window conditions
with period-by-period ranking, \QSharedRank{} with sharing, \QMulti{} with schedule
completion, and \QCombined{} with both.
Full checking of unshared QAOA candidates, initial combinations, and
fallbacks misses six conditions. Period-by-period ranking misses five more.
Combined checking recovers both groups.

At 588 evaluations, combined checking with QAOA candidates recovers
\QLowCombined{}/\PrimaryDen{}, compared with \QLowOriginal{}/\PrimaryDen{}
for period-by-period ranking and \DirectLow{}/\PrimaryDen{} for direct temporal
ranking. Common-bank frequency ranking recovers 188/196 without priority
checks and 191/196 with them. Against the latter, combined checking has
four paired gains and one loss, at 14\% in seed 601.

The ordering reverses at 618 evaluations. Both common-bank controls recover
196/196, versus 195/196 for combined checking. The latter first reaches full
recovery at 648 evaluations, and direct temporal ranking at 738.
Against period-by-period ranking, combined checking has one worse objective pair
at each of the first two quotas and none later.
In NW1, all QAOA treatments and frequency controls recover 42/42 at every quota.

All five classical methods match QAOA's combined-checking curves and reach
196/196 at 648 evaluations. Final-quota objectives and evaluated-set
infeasibility agree in all 504 paired window conditions.
Common-bank priority-control curves also coincide across all six candidate generators.
These results support schedule-based checking for \rev{either candidate source}.

\subsection{Activation requirements across windows}
At \(\eta=0.13,S=4\), the primary reference in Fig.~\ref{fig:activation}
uses counts \((3,3,3)\),
\(A=\ExampleA{},H=\ExampleH{}\), and \(G=\ExampleG{}\).
Without switching, the reference requires \(A=\ExampleStaticA{}\) device-periods.
The switching schedule uses five distinct devices, so replacements reduce
device-periods rather than installed-device count.
Lines join tested requirements, and crosses mark infeasibility.
NW1's feasible curves coincide.

\rev{The maximum \(G\) at each fixed count helps explain} this activation requirement.
At \(S=4\), counts two, three, and four attain
\(\RPrimaryKtwoSfour\%\), \(\RPrimaryKthreeSfour\%\), and
\(\RPrimaryKfourSfour\%\), respectively.
Three is therefore the smallest evaluated fixed count that reaches 13\%
in the primary window\rev{, consistent with the reference's three devices per period}. In NW1, four devices reach only
\(\RNWOneKfourSfour\%\), whereas all 16 reach \(\RNWOneKsixteenSfour\%\).
No evaluated sequence reaches 13\% there.
Counts 5--15 remain untested, so all-active failure does not establish their infeasibility.

\subsection{Detection uncertainty and attack scope}

Nonlinear WLS trials assess whether recovered schedules meet the detection
requirements. They test final-budget schedules at 13\% (primary) and 10\% (NW1)
across both caps and seven seeds.
Methods comprise all combined candidate generators, QAOA with period-by-period ranking,
references, and no-MTD/all-active controls.
The quantum-assisted sequence in Fig.~\ref{fig:operation} was fixed using
combined checking (seed 101) before detector outcomes.
The \FreshAliases{} records reuse \FreshStates{} period-combination cases
(21 primary, nine NW1).
Each case has 5000 independent calibration trials and 5000 tests per attack
and clean condition.
WLS uses solved-state initialization, tolerance \(10^{-8}\), cap 1000,
and the source pre-update residual. All \FreshFits{} fits \rev{succeed}.
Shared window/period noise draws make
schedule records dependent.

Both detectors use strict alarms \(J>\tau\) without rescheduling.
The primary threshold is nominal, whereas the secondary uses each case's
4751st clean-calibration order statistic. Calibration and test draws are disjoint,
but achieved false-alarm rates can differ.
Simultaneous 95\% Bonferroni-Clopper-Pearson intervals cover
\FreshFamily{} marginal probabilities across both detectors.
Bounds \(L_{t,s},U_{t,s}\) give the schedule-minimum interval
\([\min_{t,s}L_{t,s},\min_{t,s}U_{t,s}]\).
Confirmation requires the interval's lower endpoint to reach the requirement.

In Fig.~\ref{fig:operation}, the primary reference and quantum-assisted schedules
have equal \((A,H)\), but detection minima of \FreshRefG{}\% and
\FreshQG{}\%, with intervals \([\FreshRefLo{},\FreshRefHi{}]\%\) and
\([\FreshQLo{},\FreshQHi{}]\%\). Neither lower bound reaches 13\%.
NW1's reference interval \([\FreshNWLo{},\FreshNWHi{}]\%\) crosses 10\%.
Under the clean-calibrated detector, the reference intervals are
\([\FreshRefSecondaryLo{},\FreshRefSecondaryHi{}]\%\) (primary)
and \([\FreshNWSecondaryLo{},\FreshNWSecondaryHi{}]\%\) (NW1).
Neither detector confirms the model-based requirements.
The 13\% line is an illustrative scheduling requirement, not a protection standard.

For the nominal detector, the largest model-empirical point difference
\(|\widehat P_D-P_D^{\rm model}|\) is
\FreshModelError{} percentage points, not an error upper bound.
Simultaneous intervals support the three-percentage-point model tolerance for
106/120 attacked conditions. The other \rev{14} remain inconclusive.
Nominal false-alarm fractions are \FreshFprMin{}\%--\FreshFprMax{}\%.
All 30 clean intervals meet the two-percentage-point tolerance around 5\%.
The primary reference's upper simultaneous bound for minimum
detection probability is below the all-active control's lower bound.

Attack severity also matters. Boundary constraints keep bus 98's constructed
voltage within \(10^{-5}\) p.u. of nominal, without bounding regional
attacks, estimated post-MTD states, or control-induced damage.
Low detection therefore does not establish failure against severe undervoltage.
Fixed attack increments, known contexts, and solved-state
initialization restrict the conclusions.

\section{Conclusion}

Switching limits make complete schedules the relevant basis for allocating
physical checks in quantum-assisted D-FACTS scheduling.
Within the evaluated set, fixed-candidate comparisons show improved final-budget
recovery over period-by-period ranking for every tested candidate generator in the primary window.
Equal recovery curves for quantum and classical candidates under combined checking
support using the same allocation policy with either source.
However, common-bank frequency ranking reaches the reference objectives earlier at an intermediate
budget, so the policy must be chosen for the available evaluation allowance.
Operational use further requires detection validation under the intended attack
and noise conditions. The next step is to test schedule-based allocation on
larger candidate sets at the same physical-evaluation budget.

\section*{Acknowledgments}

\looseness=-1
This material is based upon work supported by the US National Science
Foundation under Grant No.~2146156. The authors used AI tools to polish the writing.

\bibliographystyle{IEEEtran}
\bibliography{references_r52,quantum_references_snapshot}

@article{Wang2020XYMixers,
  author  = {Zhihui Wang and Nicholas C. Rubin and Jason M. Dominy and Eleanor G. Rieffel},
  title   = {{XY} Mixers: Analytical and Numerical Results for the Quantum Alternating Operator Ansatz},
  journal = {\rev{Phys. Rev. A}},
  volume  = {101},
  pages   = {012320},
  year    = {2020},
  doi     = {10.1103/PhysRevA.101.012320}
}

@preamble{"\providecommand{\rev}[1]{#1}"}

@article{LiuWu2020Placement,
  author  = {Bo Liu and Hongyu Wu},
  title   = {Optimal {D-FACTS} Placement in Moving Target Defense Against False Data Injection Attacks},
  journal = {IEEE Trans. Smart Grid},
  volume  = {11},
  number  = {5},
  pages   = {4345--4357},
  year    = {2020},
  doi     = {10.1109/TSG.2020.2977207}
}

@article{LiuWu2021Systematic,
  author  = {{\rev{B.~Liu}} and {\rev{H.~Wu}}},
  title   = {Systematic Planning of Moving Target Defence for Maximising Detection Effectiveness against False Data Injection Attacks in Smart Grid},
  journal = {IET Cyber-Phys. Syst., Theory Appl.},
  volume  = {6},
  number  = {3},
  pages   = {151--163},
  year    = {2021},
  doi     = {10.1049/cps2.12012}
}

@article{Lakshminarayana2021Cost,
  author  = {Subhash Lakshminarayana and David K. Y. Yau},
  title   = {Cost-Benefit Analysis of Moving-Target Defense in Power Grids},
  journal = {IEEE Trans. Power Syst.},
  volume  = {36},
  number  = {2},
  pages   = {1152--1163},
  year    = {2021},
  doi     = {10.1109/TPWRS.2020.3010365}
}

@article{Lakshminarayana2021Game,
  author  = {Subhash Lakshminarayana and E. Veronica Belmega and H. Vincent Poor},
  title   = {Moving-Target Defense against Cyber-Physical Attacks in Power Grids via Game Theory},
  journal = {IEEE Trans. Smart Grid},
  volume  = {12},
  number  = {6},
  pages   = {5244--5257},
  year    = {2021},
  doi     = {10.1109/TSG.2021.3095083}
}

@article{Wang2023MMTD,
  author  = {Jiazhou Wang and Jue Tian and Yang Liu and Dong Yang and Ting Liu},
  title   = {{MMTD}: Multistage Moving Target Defense for Security-Enhanced {D-FACTS} Operation},
  journal = {IEEE Internet Things J.},
  volume  = {10},
  number  = {14},
  pages   = {12234--12247},
  year    = {2023},
  doi     = {10.1109/JIOT.2023.3245628}
}

@article{Zhang2022Voltage,
  author  = {Hang Zhang and Bo Liu and Xuebo Liu and Anil Pahwa and Hongyu Wu},
  title   = {Voltage Stability Constrained Moving Target Defense Against Net Load Redistribution Attacks},
  journal = {IEEE Trans. Smart Grid},
  volume  = {13},
  number  = {5},
  pages   = {3748--3759},
  year    = {2022},
  doi     = {10.1109/TSG.2022.3170839}
}

@article{Xu2023Robust,
  author  = {Wangkun Xu and Imad M. Jaimoukha and Fei Teng},
  title   = {Robust Moving Target Defence against False Data Injection Attacks in Power Grids},
  journal = {\rev{IEEE Trans. Inf. Forensics Security}},
  volume  = {18},
  pages   = {29--40},
  year    = {2023},
  doi     = {10.1109/TIFS.2022.3210864}
}

@article{Xu2023EventTriggered,
  author  = {Wangkun Xu and Martin Higgins and Jianhong Wang and Imad M. Jaimoukha and Fei Teng},
  title   = {Blending Data and Physics Against False Data Injection Attack: An Event-Triggered Moving Target Defence Approach},
  journal = {\rev{IEEE Trans. Smart Grid}},
  volume  = {14},
  number  = {4},
  pages   = {3176--3188},
  year    = {2023},
  doi     = {10.1109/TSG.2022.3231728}
}

@article{Jiang2026PACE,
  author  = {Yuqi Jiang and Zhiding Liang and Qiang Guan and Yan Li and Ganesh Kumar Venayagamoorthy},
  title   = {{PACE-QAOA}: Physics-Constrained Quantum Optimization for Qubit-Efficient Power System Islanding},
  journal = {\rev{arXiv:2608.02789}},
  year    = {2026},
  doi     = {10.48550/arXiv.2608.02789}
}

@article{Hadfield2019Alternating,
  author  = {Stuart Hadfield and Zhihui Wang and Bryan O'Gorman and Eleanor G. Rieffel and Davide Venturelli and Rupak Biswas},
  title   = {From the Quantum Approximate Optimization Algorithm to a Quantum Alternating Operator Ansatz},
  journal = {Algorithms},
  volume  = {12},
  number  = {2},
  pages   = {34},
  year    = {2019},
  doi     = {10.3390/a12020034}
}

@article{Jing2023DataDriven,
  author  = {Hang Jing and Ye Wang and Yan Li},
  title   = {Data-Driven Quantum Approximate Optimization Algorithm for Power Systems},
  journal = {\rev{Commun. Eng.}},
  volume  = {2},
  pages   = {12},
  year    = {2023},
  doi     = {10.1038/s44172-023-00061-8}
}

@IEEEtranBSTCTL{IEEEexample:BSTcontrol,
  CTLdash_repeated_names = "no"
}
\end{document}